\documentclass[twocolumn,aps,prl,nofootinbib]{revtex4-1}
\usepackage{ascmac,amsmath,amsthm,amssymb}
\usepackage[dvipdfmx]{graphicx}
\usepackage{color}
\usepackage{float}
\usepackage{booktabs}
\usepackage{bm}
\usepackage{overpic}
\usepackage{aas_macros}
\def\ga{\mathrel{\mathchoice {\vcenter{\offinterlineskip\halign{\hfil
$\displaystyle##$\hfil\cr>\cr\sim\cr}}}
{\vcenter{\offinterlineskip\halign{\hfil$\textstyle##$\hfil\cr>\cr\sim\cr}}}
{\vcenter{\offinterlineskip\halign{\hfil$\scriptstyle##$\hfil\cr>\cr\sim\cr}}}
{\vcenter{\offinterlineskip\halign{\hfil$\scriptscriptstyle##$\hfil
\cr>\cr\sim\cr}}}}}
\def\la{\mathrel{\mathchoice {\vcenter{\offinterlineskip\halign{\hfil
$\displaystyle##$\hfil\cr<\cr\sim\cr}}}
{\vcenter{\offinterlineskip\halign{\hfil$\textstyle##$\hfil\cr<\cr\sim\cr}}}
{\vcenter{\offinterlineskip\halign{\hfil$\scriptstyle##$\hfil\cr<\cr\sim\cr}}}
{\vcenter{\offinterlineskip\halign{\hfil$\scriptscriptstyle##$\hfil
\cr<\cr\sim\cr}}}}}
\begin{document}
\title{Lagrangian cosmological perturbation theory at shell-crossing}

\author{Shohei~Saga}
\affiliation{Yukawa Institute for Theoretical Physics, Kyoto University, Kyoto 606-8502, Japan}

\author{Atsushi~Taruya}
\affiliation{Center for Gravitational Physics, Yukawa Institute for Theoretical Physics, Kyoto University, Kyoto 606-8502, Japan}
\affiliation{Kavli Institute for the Physics and Mathematics of the Universe (WPI), Todai institute
for Advanced Study, University of Tokyo, Kashiwa, Chiba 277-8568, Japan}

\author{St\'ephane Colombi}
\affiliation{Institut d'Astrophysique de Paris, CNRS UMR 7095 and Sorbonne Universit\'es, 98bis bd Arago, F-75014 Paris, France}

\date{\today}
\begin{abstract}
We consider the growth of primordial dark matter halos seeded by three crossed initial sine waves of various amplitudes. Using a Lagrangian treatment of cosmological gravitational dynamics, we examine the convergence properties of a high-order perturbative expansion in the vicinity of shell-crossing, by comparing the analytical results with state-of-the-art high resolution Vlasov-Poisson simulations. Based on a quantitative exploration of parameter space, we study explicitly for the first time the convergence speed of the perturbative series, and find, in agreement with intuition, that it slows down when going from quasi one-dimensional initial conditions (one sine wave dominating) to quasi triaxial symmetry (three sine waves with same amplitude). In most cases, the system structure at collapse time is, as expected, very similar to what is obtained with simple one-dimensional dynamics, except in the quasi-triaxial regime, where the phase-space sheet presents a velocity spike. In all cases, the perturbative series exhibits a generic convergence behavior as fast as an exponential of a power-law of the order of the expansion, allowing one to numerically extrapolate it to infinite order. The results of such an extrapolation agree remarkably well with the simulations, even at shell-crossing.
\end{abstract}

\preprint{YITP-18-49}
\maketitle
{\it Introduction.---}%
The observed large-scale structures of the Universe are thought to be mainly the result of the evolution through gravitational instability of small initial fluctuations in the matter distribution, with a dominant component given by collisionless dark matter. In the concordance model, Cold Dark Matter (CDM) \cite{1982ApJ...263L...1P,1984ApJ...277..470P,1984Natur.311..517B}, the dark matter particles form, at the macroscopic level, a smooth distribution with a virtually null initial local velocity dispersion. This means that dark matter dynamics follow Vlasov-Poisson equations and that the dark matter distribution can be represented as a three-dimensional sheet evolving in six-dimensional phase-space.

In the standard CDM picture, the phase-space sheet is initially perturbed by small Gaussian random fluctuations in the density distribution, with a cut-off scale related to the mass of the dark matter particle \cite{1982ApJ...263L...1P,2003PhRvD..68j3003B}. The first nonlinear structures form at shell-crossing, i.e. in regions of space where the phase-space sheet first self-intersects. Primordial dark matter halos with a power-law density profile of logarithmic slope around $-1.5$ \cite{2005Natur.433..389D,2014ApJ...788...27I,2017MNRAS.471.4687A,2018MNRAS.473.4339O} grow around these initial singularities through violent relaxation and then merge together hierarchically to form larger halos with universal but different properties \cite{1996ApJ...462..563N,1997ApJ...490..493N,2010MNRAS.402...21N,2011ASL.....4..297D,1997ApJ...477L...9F}.
Due to the complexity of post-collapse dynamics, the actual processes leading to the shape of a primordial or an evolved halo remain a subject of debate. Indeed, there is no exact analytical theory to predict the results of numerical simulations, although many approaches have been proposed to tackle the problem, relying on e.g. self-similarity \cite{1984ApJ...281....1F,1985ApJS...58...39B,2013MNRAS.428..340A} or entropy maximization \cite{1967MNRAS.136..101L,2010ApJ...722..851H,2013MNRAS.432.3161C,2013MNRAS.430..121P}. In this debate, the detailed knowledge of the structure of primordial dark matter protohalos prior to shell-crossing seems essential and even this remains a challenge in the general case. 

To describe gravitational dynamics before shell-crossing, it is however possible to employ perturbation theory (PT) as long as fluctuations in the density field remain small \cite{2002PhR...367....1B}. With Lagrangian perturbation theory (LPT) \cite{1970A&A.....5...84Z,1989RvMP...61..185S,1992ApJ...394L...5B,1992MNRAS.254..729B,1993MNRAS.264..375B,1995A&A...296..575B,1994ApJ...427...51B}, which uses the displacement field as a small parameter in the expansion of the equations of motion, one can  rather realistically follow the evolution of a system in the nonlinear regime up to shell-crossing or even slightly beyond. Zel'dovich approximation \cite{1970A&A.....5...84Z}, which corresponds to linear order, has been widely used in the literature. It already gives the exact solution until shell-crossing in the one-dimensional case \cite{1969JETP...30..512N} and also provides, in higher dimension, a firm framework to study the families of singularities that first form at shell-crossing \cite{1982GApFD..20..111A,2014MNRAS.437.3442H,2017arXiv170309598F}. In general, higher order PT is required to have a sufficiently accurate description of pre-collapse dynamics \cite{1991ApJ...382..377M,1997A&A...318....1B,1998ApJ...498...48Y} and this obviously depends on the nature of initial conditions. While the radius of convergence of the perturbative series has been studied for LPT \cite{2014JFM...749..404Z,2015MNRAS.452.1421R}, the speed of such convergence has been little investigated and remains an important question.

In this Letter, we compare predictions of LPT up to tenth order to state-of-the-art Vlasov-Poisson simulations, for primordial halos initially seeded by three crossed sine waves. Such configurations are still generic, as they can be assimilated to high peaks of a smooth random Gaussian field \cite{1986ApJ...304...15B}. Varying the amplitudes of the sine waves will allow us to span a realistic range of initial configurations, from quasi-unidimensional to quasi-triaxial. We analyze the phase-space structure of such systems at shell-crossing and study the speed of convergence of LPT according to initial set-up, including extrapolation to infinite order.

{\it Setup.---}%
In the presence of gravity, the Lagrangian equation of motion of a matter element in the expanding Universe is given by
\begin{align}
\frac{{\rm d}^2\bm{x}}{{\rm d}t^2} + 2H \frac{{\rm d}\bm{x}}{{\rm d}t} &= -\frac{1}{a^{2}}\bm{\nabla}_{x} \phi(\bm{x}) , 
\label{eq:basic_EoM}
\end{align}
where $\bm{x}$ is the comoving position, $\phi$ the Newton potential obtained by solving Poisson equation,
\begin{align}
\bm{\nabla}^{2}_{x}\phi(\bm{x}) &= 4\pi G \bar{\rho}_{\rm m} a^{2} \delta(\bm{x}) , 
\label{eq:Poisson}
\end{align}
and where the quantities $a$, $H=a^{-1} {\rm d}a/{\rm d}t$, $\bar{\rho}_{\rm m}$, and $\delta$ correspond to the scale factor of the Universe, Hubble parameter, background mass density and matter density contrast, respectively. In this framework, the velocity of each mass element is given by $\bm{v} = a\,{\rm d}\bm{x}/{\rm d}t$. 

Here, we use a Lagrangian approach i.e. we follow motion as a function of initial position, the Lagrangian coordinate $\bm{q}$. The subsequent evolution of the position is expressed as $\bm{x}(\bm{q},t)=\bm{q}+ \bm{\Psi}(\bm{q},t)$, where $\bm{\Psi}$ represents the displacement from initial position. Then, the velocity field is expressed as $\bm{v}(\bm{q},t)=a(t) {\rm d}\bm{\Psi}/{\rm d}t$, and, in the single flow regime, mass conservation implies $d^3\bm{q}=\{1+\delta(\bm{x})\}\,d^3\bm{x}$, which leads to $1 + \delta(\bm{x}) = 1/J$ with $J=\det{\left( \delta_{ij} + \Psi_{i,j}\right)}$. These last equations are valid until shell-crossing time $t_{\rm sc}$, that is the first occurrence of $J=0$.

We start with initial conditions given by three crossed sine waves in a periodic box covering the interval $[-L/2, L/2[$:
\begin{align}
\Psi_A^{\rm ini}(\bm{q}, t_{\rm ini}) =\frac{L}{2\pi}\,D_+(t_{\rm ini})\,\epsilon_{A} \sin\left( \frac{2\pi}{L}q_{A}\right),\,\,(A=x,y,z), 
\label{eq:init_psi}
\end{align}
where $D_+$ is the linear growth factor.
The initial time, $t_{\rm ini}$, and parameters, $\epsilon_A < 0$, $|\epsilon_x| \geq |\epsilon_y| \geq |\epsilon_z|$, are chosen such that $D_+(t_{\rm ini})|\epsilon_A|\ll1$, so only e.g. the ratios $\epsilon_y/\epsilon_x$ and $\epsilon_z/\epsilon_x$ are relevant. Our analytical investigations will cover the full range of values of $\epsilon_{y,z}/\epsilon_x$, while the simulations will address three types of configurations:
quasi one-dimensional with $|\epsilon_x| \gg |\epsilon_{y,z}|$, triaxial asymmetric with $|\epsilon_x| > |\epsilon_y| > |\epsilon_z|$, and triaxial symmetric with $|\epsilon_x|=|\epsilon_y|=|\epsilon_z|$, denoted by Q1D-S, ASY-S and SYM-S, respectively.

{\it Lagrangian Perturbation Theory.---}%
In LPT, the displacement field $\bm{\Psi}$ is the fundamental building block which is considered as a small quantity. As long as function $\bm{x}(\bm{q})$ is a single-valued function of $\bm{q}$, there is no singularity in the density field and a systematic perturbative expansion is possible, $\bm{\Psi} = \sum_{n = 1}^{\infty}\bm{\Psi}^{(n)}$. The evolution equation of the displacement field at each order is obtained from Eq.~(\ref{eq:basic_EoM}). Taking the divergence and the curl of this equation in Eulerian coordinates and rewriting the expressions in Lagrangian coordinates, a set of recurrence relations is obtained by substituting the perturbative expansion into Eq.~(\ref{eq:basic_EoM}) \cite{2012JCAP...12..004R,2014JFM...749..404Z,2015MNRAS.452.1421R,2015PhRvD..92b3534M}:
\begin{align}
&\left( \hat{\mathcal{T}} - \frac{3}{2}\right)\Psi^{(n)}_{k,k}
=
-\varepsilon_{ijk}\varepsilon_{ipq}
\sum_{n=a+b}
\Psi^{(a)}_{j,p}\left( \hat{\mathcal{T}} - \frac{3}{4}\right) \Psi^{(b)}_{k,q} \notag \\
&
-\frac{1}{2}\varepsilon_{ijk}\varepsilon_{pqr} \sum_{n = a+b+c}\Psi^{(a)}_{i,p}\Psi^{(b)}_{j,q}\left( \hat{\mathcal{T}} - \frac{1}{2}\right)\Psi^{(c)}_{k, r} , \label{eq:longitudinal2}
\end{align}
for the longitudinal part, and
\begin{align}
\varepsilon_{ijk}\hat{\mathcal{T}} \Psi^{(n)}_{j,k}
= -\varepsilon_{ijk}\sum_{n = a + b}\Psi^{(a)}_{p,j} \hat{\mathcal{T}} \Psi^{(b)}_{p,k}, 
\label{eq:transverse2}
\end{align}
for the transverse part. Here, $\varepsilon_{ijk}$ is the Levi-Civita symbol, $\Psi_{i,j}\equiv\partial\Psi_i/\partial q_j$ and $\hat{\mathcal{T}}$ stands for a differential operator,
$\hat{\mathcal{T}}\equiv (\partial^2/\partial \tau^2)+\frac{1}{2}(\partial/\partial \tau)$, where $\tau \equiv \ln{D_{+}(t)}$.
Thus, according to the initial conditions given by Eq.~(\ref{eq:init_psi}), one builds up from Eqs.~(\ref{eq:longitudinal2}) and (\ref{eq:transverse2}) the perturbative solutions for the two kinds of derivatives $\nabla\cdot\bm{\Psi}^{(n)}$ and $\nabla\times\bm{\Psi}^{(n)}$. Then, the displacement field can be consistently reconstructed from these derivatives. This last step is non-trivial and generally involves intricate calculations. However, thanks to the trigonometric polynomial nature of the initial conditions we consider, $\bm{\Psi}^{(n)}$ is expressed likewise after simple algebraic manipulations.

Since the initial set-up assumes a small amplitude of the fluctuations, $D_+(t_{\rm ini})|\epsilon_A|\ll1$, and we consider time sufficiently close to collapse, $D_+(t)|\epsilon_A| \sim 1$, taking only the fastest growing-mode provides an accurate description of the dynamics.
With this additional simplification, we perform the perturbative calculation in two different ways, as follows. The first approach, quite standard, consists in using Eq.~(\ref{eq:init_psi}) as the first-order solution, i.e., $\bm{\Psi}^{(1)}(\bm{q},t)=\bm{\Psi}^{\rm ini}(\bm{q},t)$, and develop higher-order calculations (LPT) up to tenth order. The second approach, that we denote by Q1D, assumes quasi one-dimensional dynamics, following the footsteps of \cite{2017MNRAS.471..671R}, i.e. $|\epsilon_x| \gg |\epsilon_{y,z}|$. In this case, one takes the exact solution of the one-dimensional problem along $x$-axis given by Zel'dovich approximation as the unperturbed zeroth-order state: $\Psi^{(0)}_A(\bm{q},t) = \Psi^{{\rm ini}}_{x}(q_{x},t) \,\delta_{A,x}$, $A=x,y,z$, with $\delta_{A,B}$ being the Kronecker delta. Transverse fluctuations are considered as small first-order perturbations to this set-up, $\Psi^{(1)}_A(\bm{q},t) = \Psi^{\rm ini}_{y}(q_{y}, t)\,\delta_{A,y}+ \Psi^{\rm ini}_{z}(q_{z},t)\,\delta_{A,z}$. The perturbative expansion is then performed by keeping terms proportional to $\epsilon_y^i$ and $\epsilon_z^j$ up to second order, $i+j=2$ (so in this sense we go one order beyond \cite{2017MNRAS.471..671R}), while keeping terms proportional to $\epsilon_x^k$ up to tenth order, $k=10$. 

{\it Vlasov simulations and phase-space structure.---}%
To quantitatively investigate the dynamics of our system up to shell-crossing, we perform high resolution simulations with the state-of-the-art Vlasov-Poisson code \texttt{ColDICE}~\cite{2016JCoPh.321..644S}. This code uses a tessellation, i.e. tetrahedra, to represent the phase-space sheet. The vertices of the tessellation form initially a homogeneous mesh of resolution $n_{\rm s}$ (which corresponds to $6n_{\rm s}^3$ simplices) and then follow Lagrangian equations of motion, Eqs.~(\ref{eq:basic_EoM}) and (\ref{eq:Poisson}). Poisson equation is solved by Fast-Fourier-Transform on a regular cartesian grid of resolution $n_{\rm g}$, after deposit of the phase-space sheet on the grid using the method of Franklin and Kankanhalli generalized to linear order \cite{FRANKLIN1983327,Franklin:1987:PPC:41958.41969,10.1007/3-540-56869-7_27}. A number of simulations, as listed in Table~\ref{tab:tab}, were performed assuming Einstein-de Sitter cosmology.
\begin{table}
\begin{tabular}{lcccc}
\hline
Designation & $D_+(t_{\rm ini}) |\epsilon_x|$ & $(\epsilon_y/\epsilon_x,\epsilon_z/\epsilon_x)$ & $n_{\rm s}$ & $n_{\rm g}$ \\
\hline
Q1D-S & 0.012 & (0.167,0.125) & 256 & 512 \\
ASY-Sa & 0.012 & (0.625,0.5) & 512 & 512 \\
ASY-Sb & 0.012 & (0.75,0.5) & 512 & 512 \\
ASY-SbHR & 0.012 & (0.75,0.5) & 512 & 1024 \\
ASY-Sc & 0.012 & (0.875,0.5) & 512 & 512 \\
SYM-S & 0.009 & (1,1) & 512 & 512 
\end{tabular}
\caption[]{Parameters of the runs performed with \texttt{ColDICE}.}
\label{tab:tab}
\end{table}

\begin{figure}[t]
\begin{center}
\includegraphics[width=0.37\textwidth]{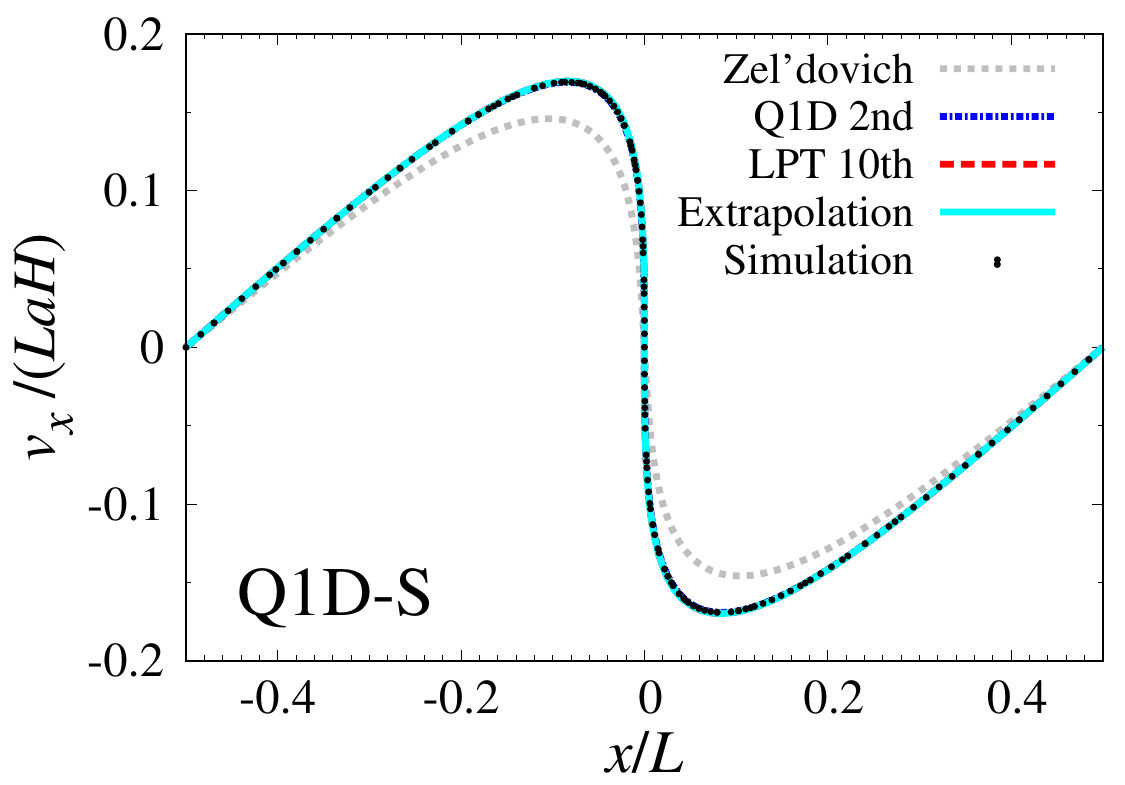}
\includegraphics[width=0.37\textwidth]{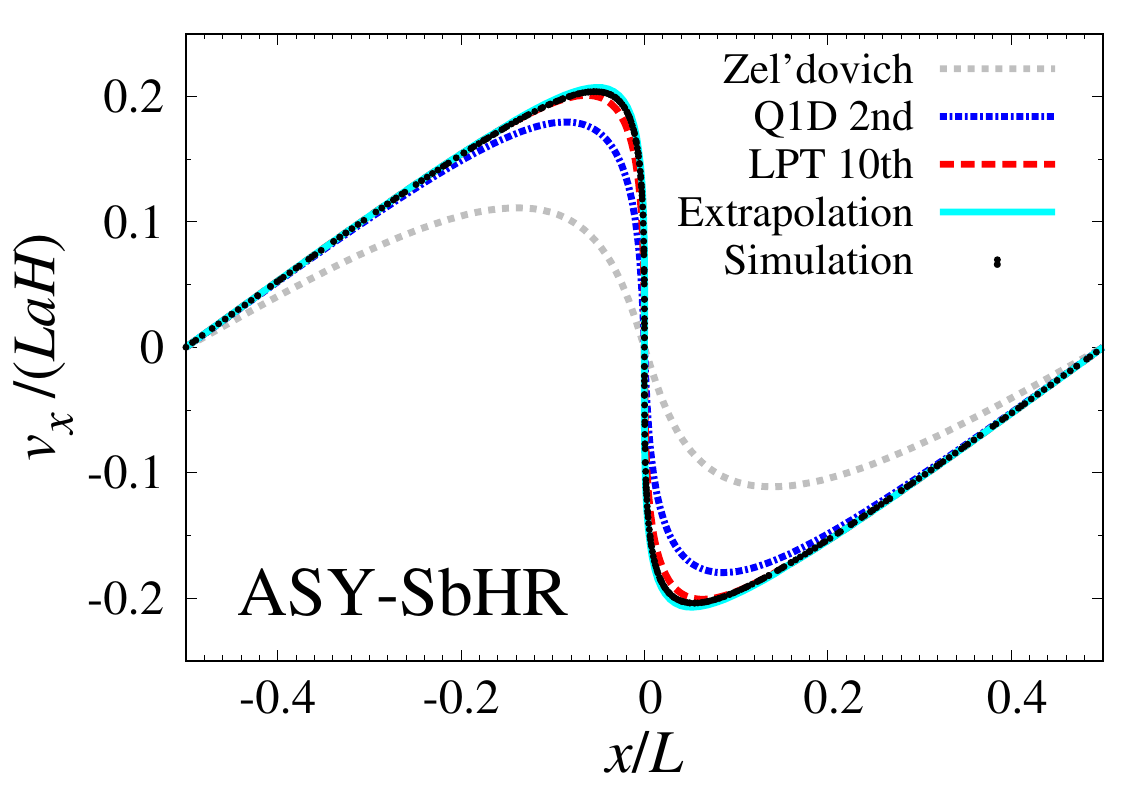}
\includegraphics[width=0.37\textwidth]{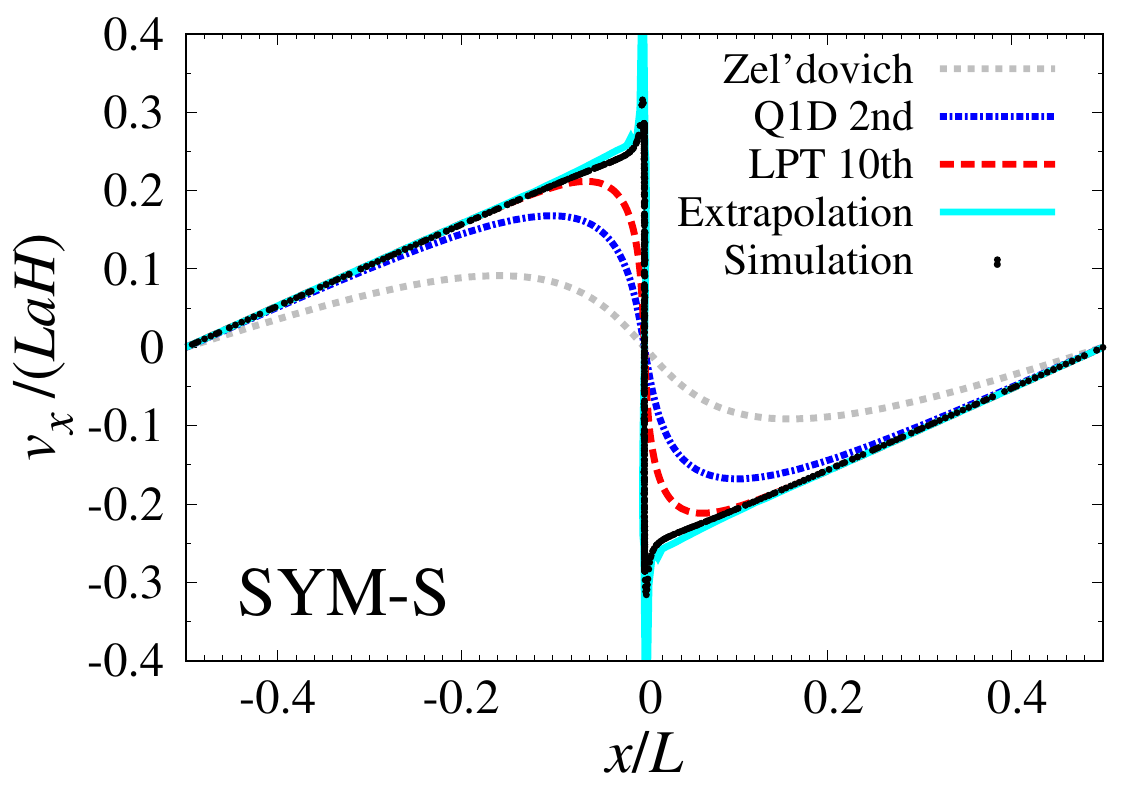}
\end{center}
\caption{%
Phase-space structure at collapse time $t_{\rm sc}$: the intersection of the phase-space sheet with $y=z=0$ hyper-plane is displayed in $(x,v_{x})$ subspace for runs Q1D-S, ASY-SbHR and SYM-S, from top to bottom.
The simulation (black points) is compared to Zel'dovich approximation (grey dots), Q1D (blue dot-dashes) and LPT up to tenth order (red dashes), as well as the extrapolated method (cyan curve).
On the top panel, all the curves superpose to each other except for the grey dots.
From top to bottom, shell-crossing time $t_{\rm sc}$ corresponds to 
$|\epsilon_{x}|D_+(t_{\rm sc}) \approx 0.912$, 
$0.696$ and 
$0.576$, respectively. 
Note that, in middle panel, the lower-resolution simulation, ASY-Sb, would give undistinguishable results from ASY-SbHR, showing that using a $512^3$ grid to solve Poisson equation is sufficient to achieve convergence of the numerical experiments. 
}
\label{fig: result x-vx}
\end{figure}


Fig.~\ref{fig: result x-vx} shows representative examples of phase-space portraits at shell-crossing time. As the ratios $\epsilon_{y,z}/\epsilon_x$ increase, Zel'dovich approximation, which is exact in the strictly one-dimensional case $\epsilon_{y,z}/\epsilon_x=0$, starts to deviate significantly from the simulation, as expected. The Q1D prediction provides a substantial improvement, with an excellent match of the simulation measurements for $\epsilon_{y,z}/\epsilon_x\ll1$ (top panel). Still, it cannot catch up the shell-crossing structure when ratios $\epsilon_{y,z}/\epsilon_x$ approach unity (middle and bottom panels). However, with a systematic calculation of all the contributions up to tenth order, LPT prediction improves considerably and accurately reproduces the shell-crossing structure seen in the simulations for most of the parameter space, except in the vicinity of $(\epsilon_y/\epsilon_x,\epsilon_z/\epsilon_x)=(1,1)$ (bottom panel).

In the $\epsilon_{y,z}/\epsilon_x=1$ case, the phase-space structure is highly stretched along velocity axis, which reflects a noticeable acceleration of the inward mass flow near $\bm{x}=0$, similarly as in spherical collapse \cite{1984ApJ...281....1F,1985ApJS...58...39B}. This highly contrasted dynamical behavior slows down LPT convergence near $\epsilon_{y,z}/\epsilon_x=1$ and even the tenth-order calculation is insufficient (see also \cite{2017arXiv171201878R} for a discussion about the spherically symmetric case). 

However, by studying the sequence of LPT predictions as a function of order $n$ up to $n=10$, it is possible to extrapolate the asymptotic convergence at $n\to\infty$. Indeed, we find that the position of matter elements at collapse time calculated at $n{\rm th}$ order with LPT, $\bm{x}_{\rm sc}(\bm{q},n) =[x_{\rm sc}(\bm{q},n),y_{\rm sc}(\bm{q},n),z_{\rm sc}(\bm{q},n)]$, is well described by the following fitting form: 
\begin{align}
A_{\rm sc}(\bm{q},n) = \alpha_A(\bm{q}) + \frac{1}{b_A(\bm{q}) + c_A(\bm{q}) \exp\left[ d_{A}(\bm{q})\,n^{e_{A}(\bm{q})}\right]}, 
\label{eq:fitting_x}
\end{align}
with $A=x,y,z$ and where $\alpha_A$, $b_A$, $c_A$, $d_A$ and $e_A$ are fitting parameters which depend both on initial conditions and Lagrangian position $\bm{q}$. Taking the limit $n\to\infty$ gives the extrapolated result at infinite order, $\bm{x}_{\rm sc}(\bm{q}, n \rightarrow \infty)\to [\alpha_x(\bm{q}),\alpha_y(\bm{q}),\alpha_z(\bm{q})]$. The same treatment can be applied to compute shell-crossing time $t_{\rm sc}(n \rightarrow \infty)$ used as the output time for the simulations in Fig.~\ref{fig: result x-vx}, while the extrapolated velocity, $\bm{v}_{\rm sc}(\bm{q},n \rightarrow \infty)$, is obtained by finite time differences on the position. Note that form (\ref{eq:fitting_x}) is not the unique choice, but using the exponential of a power-law might be the only way to match the convergence speed of LPT at large $n$.

Examination of Fig.~\ref{fig: result x-vx} shows that the result of this procedure (cyan curves) reproduces very well simulation measurements, even the spiky structure in the $\epsilon_{y,z}/\epsilon_x=1$ case. Disagreement is at worse a few percent when $\epsilon_{y,z}/\epsilon_x<1$. Even if there are still some small mismatches, partly attributable to a small desynchronization due to the finite time step in the simulations, the extrapolation based on Eq.~(\ref{eq:fitting_x}) is unquestionably successful and can be used to perform quantitative predictions over the entire parameter space covered by $(\epsilon_y/\epsilon_x,\epsilon_z/\epsilon_x)$. 

{

{\it Exploration of parameter space.---}%
Making use of the generic behavior described in Eq.~(\ref{eq:fitting_x}), we can explore convergence properties of LPT as well as the global parameter dependence of the shell-crossing structure. In agreement with intuition, LPT convergence worsens when going from quasi one-dimensional to quasi-triaxial symmetric, as illustrated by top panel of Fig.~\ref{fig:eps_plane} for shell-crossing time. In general, few percent accuracy requires high order LPT. For instance, 3 percent accuracy examined in Fig.~\ref{fig:eps_plane} can be achieved at third order only for $(\epsilon_y+\epsilon_z)/\epsilon_x \la 0.5$, which represents merely one-eighth of the total parameter space, and probing half the parameter space would require seventh order.

Bottom panel of Fig.~\ref{fig:eps_plane} focuses on the maximum velocity $v_{x}/(LaH)$. As expected, for the parameters probed by our runs, the theoretical predictions given by the black dots are found to be in good agreement with the simulations. What is more striking is the sudden augmentation of the maximum velocity in the vicinity of $(\epsilon_y/\epsilon_x,\epsilon_z/\epsilon_x)=(1,1)$. This sudden change is associated to a drastic variation of the phase-space structure at shell-crossing, as illustrated by Fig.~\ref{fig:varying_eps}, where we consider the case $\epsilon_z/\epsilon_x=1$ and values of $\epsilon_y/\epsilon_x$ increasing from $0.85$ to unity. As seen in this figure, the cross-section of the phase-space sheet changes drastically from a smooth ``S'' shape, which is the normal behavior for most of values of the ratios $\epsilon_{y,z}/\epsilon_x$, to a spiky structure when both these ratios approach unity, ${\rm min}(\epsilon_{y,z}/\epsilon_x) \ga 0.9$. While the presence of a spiky structure in the quasi-triaxial symmetric case can be expected, as it is found in spherical collapse, the way it appears in parameter space remains non trivial.

\begin{figure}[t]
\begin{center}
\includegraphics[width=0.48\textwidth]{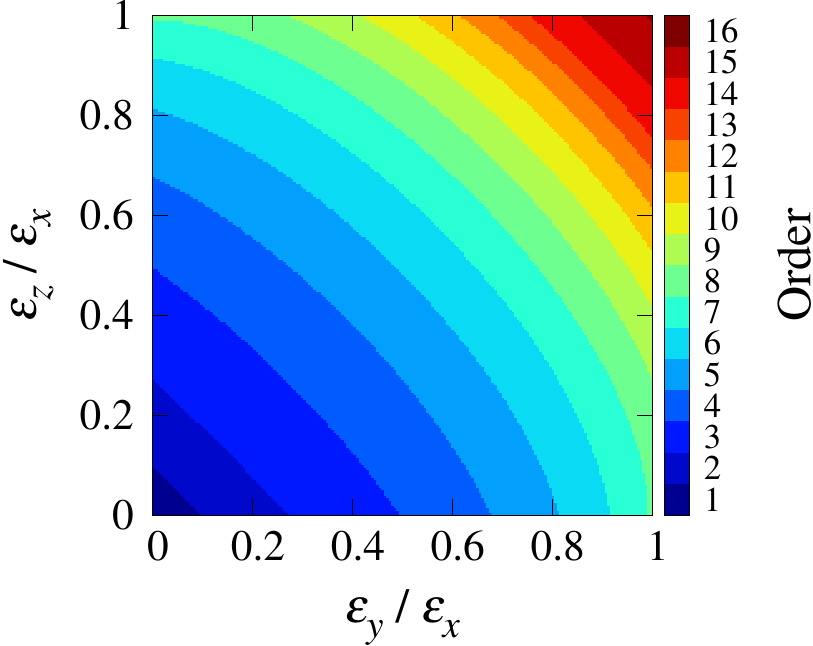}
\includegraphics[width=0.5\textwidth]{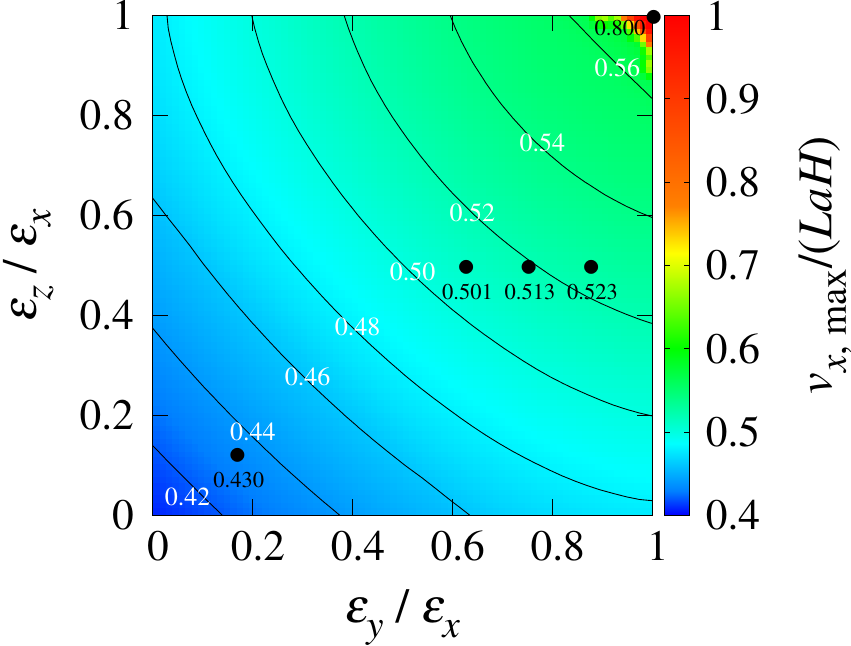}
\end{center}
\caption{%
Exploration of parameter space. {\it Top panel:} order at which LPT convergence remains better than 3 percent when calculating shell-crossing time. {\it Botton panel:} parameter dependence of the maximum velocity $v_{x,{\rm max}}/(La H)$, which is normalized to unity for $(\epsilon_{y}/\epsilon_{x},\epsilon_{z}/\epsilon_{x}) = (1,1)$. The black dots correspond to the parameters used for our runs (Table~\ref{tab:tab}), 
along with measured values of $v_{x,{\rm max}}/(La H)$ in the simulations, to be compared to the isocontours.
}
\label{fig:eps_plane}
\end{figure}
\begin{figure}[t]
\begin{center}
\includegraphics[width=0.4\textwidth]{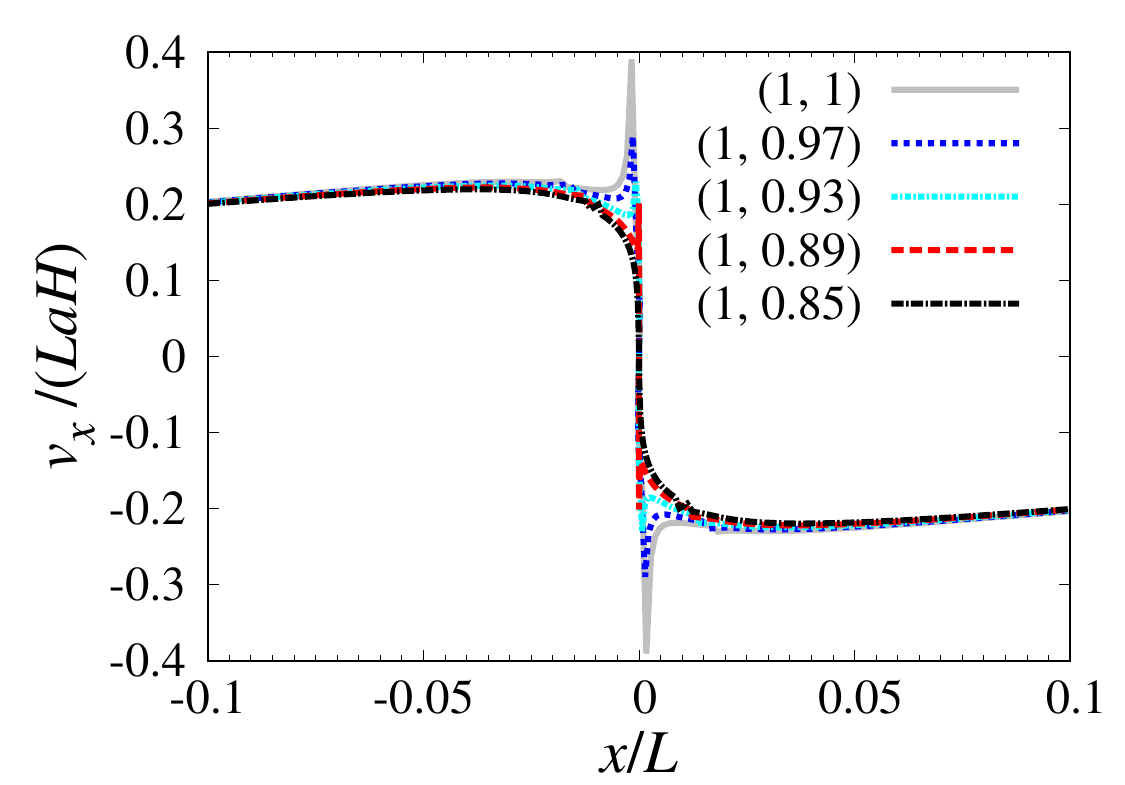}
\end{center}
\caption{%
The phase-space structure predicted by perturbation theory, similarly as in Fig.~\ref{fig: result x-vx}. Here, the results extrapolated at infinite order are shown for various values of $(\epsilon_{y}/\epsilon_{x},\epsilon_{z}/\epsilon_{x})$, ranging from $(1,0.85)$ to $(1,1)$.
}
\label{fig:varying_eps}
\end{figure}

{\it Conclusion and outlook.---}%
With Lagrangian perturbation theory (LPT) extrapolated to infinite order, we found a way to describe accurately the phase-space structure of protohalos growing from three initial sine waves of various amplitudes, $\epsilon_x$, $\epsilon_y$ and $\epsilon_z$, until collapse time.
To validate the theory, we used the state-of-the-art Vlasov code~\cite{2016JCoPh.321..644S}. Based on an exploration of parameter space, we checked that convergence of the LPT series expansion slows down when going from quasi-one dimensional to triaxial symmetric initial conditions. This exploration also shows that a spiky structure in phase-space appears when approaching triaxial symmetry, ${\rm min}(\epsilon_{y,z}/\epsilon_x) \ga 0.9$. Such a spiky structure might correspond in the CDM paradigm to a population of rare halos or subhalos with particular properties and is worth being the object of future investigations.

We are confident that our results are quite generic, even if we considered a restricted class of initial conditions. Firstly, as mentioned in the introduction, three sine waves are representative, to a large extent, of high peaks of a smooth random Gaussian density field. Secondly, our LPT algorithm can be generalized to any trigonometric polynomial initial conditions, which will allow us in the near future to account for tidal effects in the dynamical process of protohalo formation.
Our accurate prediction for the collapse time may also be used in excursion set treatments to improve predictions of the halo mass function (see \cite{2018PhR...733....1D} for a comprehensive review).

Furthermore, our LPT calculations set up the framework for accurate theoretical investigations beyond collapse time. Indeed, except for rare cases such as the triaxial symmetric configuration, collapse is generally expected to produce a planar singularity (as illustrated by two top panels of Fig.~\ref{fig: result x-vx}), meaning that near collapse time, dynamics is quasi unidimensional \cite{1970A&A.....5...84Z}. Starting from the state of the system at crossing time, it is possible to generalize the post-collapse LPT formalism developed in one dimension by \cite{2015MNRAS.446.2902C,2017MNRAS.470.4858T} to the fully three-dimensional case. Indeed, one can make use of the quasi unidimensionality of the singularity at collapse time to compute the asymptotic dynamical behavior of the system shortly after it, with the proper Taylor expansions in space and time. We leave this project for future work. Still, describing the full merging history of dark matter halos will remain a challenge.

{\it Acknowledgements.---}%
This work was supported in part by JSPS Research Fellow Grant-in-Aid Number 17J10553 (SS) and MEXT/JSPS KAKENHI Grants, Numbers JP15H05889 and JP16H03977 (AT), as well as ANR grant ANR-13-MONU-0003 (SC). Numerical computation was carried out using the HPC resources of CINES (Occigen supercomputer) under the GENCI allocation 2017-A0040407568. It has also made use of the Yukawa Institute Computer Facility. 

\bibliographystyle{apsrev4-1}
\bibliography{ref}
\end{document}